\documentclass{mn2e}

 \usepackage[dvips]{epsfig}
 \usepackage{amsfonts,amsbsy,amssymb}

\title[4U 1735--44 ephemeris] 
  {The optical ephemeris and X-ray variability of 4U 1735--44 (V926 Sco)}
\author[A. B. Giles and K. M. Hill]
       {A. B. Giles$^{1,2 \bigstar}$ and K. M. Hill$^{1}$\\
      $^{1}$ School of Mathematics and Physics, University of Tasmania, 
             Private Bag 37, Hobart, Tasmania 7001, Australia \\ 
      $^{2}$ Antarctic Climate and Ecosystems CRC, University of Tasmania, 
             Private Bag 80, Hobart, Tasmania 7001, Australia  }

\date{Submitted:  5 March 2012}

\pagerange{\pageref{firstpage}--\pageref{lastpage}}
\pubyear{2012}

\begin{document}

\maketitle

\label{firstpage}

\begin{abstract}
Optical observations of the low mass X-ray binary 4U 1735--44 
were obtained during 1997 -- 2007 and combined with earlier published 
observations from 1984 -- 1993 to refine the ephemeris for the system. 
The linear fit  for the time of maximum optical light has the ephemeris 
HJD = 2445904.0494(90) + [ N $\times$ 0.19383222(29) ] 
with a value of $\chi^2 = 253.5$ for 16 dof and a scatter about phase 
zero of $\sigma$ = 0.061. The new data reconciles the discrepancy 
between the previous ephemeris and the more recent spectral ephemeris, 
based on emission from the companion star, which defined the systems 
true dynamical phase zero. The optical maximum 
for 4U 1735--44 now occurs at spectral phase 0.47 $\pm$ 0.05 and thus 
supports the classic model for an LMXB system.  Our data further supports 
the standard model in several ways. The mean optical flux shows a positive 
correlation with the {\it RXTE} ASM X-ray flux, the relative increases 
suggesting that the non-X-ray induced optical flux from 
the companion is $\lesssim 14$ percent of the total optical light from 
the system. There is no apparent trend in the optical modulation percentage 
with increasing X-ray flux. The X-ray flux for various ASM energy bands shows 
no evidence of orbital modulation, eclipses or dips when folded using 
the new ephemeris. \end{abstract}


\renewcommand{\thefootnote}{ }
\footnote{$^{\bigstar}$ E-mail:  Barry.Giles$@$utas.edu.au}

\section{Introduction}

Millisecond oscillations in the X-ray brightness of low mass X-ray 
binaries (LMXB) have been observed for a large and growing number 
of sources since the launch of the {\it Rossi X-ray Timing Explorer} 
({\it RXTE}) in late 1995. A review of those sources which exhibit nearly 
coherent oscillations during their Type I X-ray bursts, a class that 
includes 4U 1636--53, can be found in Strohmayer (2001). The larger 
group that exhibit quasi-periodic oscillations only during their 
quiescent non-bursting state, a class that includes 4U 1735--44, have 
been reviewed by Van der Klis (2000). Relatively few of the non-transient 
X-ray sources exhibiting kHz oscillations have permanent optical 
counterparts but the similar systems 4U 1735--44 and 4U 1636--53 
are ideal for studies at optical wavelengths with small telescopes. 
The many similarities of these two syatems are discussed in Casares 
et al (2006) and references therein and will not be repeated here. 
A more general summary of the many known X-ray emitting accreting 
neutron stars can be found in Watts et al. (2008). 

In 1997 we commenced a program of occasional monitoring of 4U 1735--44 
and 4U 1636--53. The initial thrust of this work was twofold in nature. 
One program was to obtain an updated ephemeris for 4U 1636--53 and to use 
this to examine the binary phases of a large sample of X-ray bursts 
observed with {\it RXTE}. The intention was to search for possible Doppler 
shift modulations in the bursts kHz oscillation frequency induced by the 
orbital motion of the neutron star. This work has been reported by Giles et 
al. (2002). The initial start on 4U 1735--44 was part of a coordinated 
program in 1997 to observe simultaneous X-ray and optical bursts but no 
X-ray burst were seen ({\it RXTE} observation 20084). In this paper we 
present our data and results for CCD optical photometry of the 
4U 1735--44 system obtained since 1997. 

\begin{table*}
\centering
\begin{minipage}{175mm}
\begin{center}
  \bf Table 1. \rm Optical observations of 4U 1735--44. \\
  \vspace*{0.125cm}
 \label{symbols}
 \begin{tabular}{@{}llrrccc}
 \hline
 Obs. &                &   HJD Start    &   HJD End      &             &  Integration   &  Number of  \\
 No.  &    Date        &   (-2450000)   &   (-2450000)   &    Filter   &  Time (s)      &  Exposures  \\
 \hline
  1   &   1997 Aug  1  &    661.92966   &    662.13406   &  {\it V\/}, {\it I\/} & 120  &     64      \\
  2   &   1997 Aug 29  &    689.89771   &    690.08805   &  {\it V\/}, {\it I\/} &  60  &    115      \\
  3   &   1999 Sep 16  &   1437.89914   &   1438.11925   &  {\it V\/}  &     150        &     93      \\
  4   &   1999 Sep 20  &   1441.89958   &   1442.11134   &  {\it V\/}  &     300        &     52      \\
  5   &   2000 May 25  &   1689.90368   &   1690.30974   &  {\it V\/}  &     300        &     92      \\
  6   &   2001 Jul  3  &   2093.89874   &   2094.32339   &  {\it V\/}  &     300        &    110      \\
  7   &   2001 Jul  4  &   2094.90880   &   2095.31425   &  {\it V\/}  &     300        &    111      \\
  8   &   2004 May  7  &   3133.09892   &   3133.34262   &  {\it V\/}  &     300        &     63      \\
  9   &   2004 May 14  &   3139.92629   &   3140.35182   &  {\it V\/}  &     300        &    115      \\
 10   &   2004 May 15  &   3141.01611   &   3141.18857   &  {\it V\/}  &     300        &     46      \\
 11   &   2005 May 12  &   3502.95212   &   3503.33362   &  {\it V\/}  &     300        &     80      \\
 12   &   2005 May 13  &   3503.92808   &   3504.31953   &  {\it V\/}  &     300        &    106      \\
 13   &   2007 Apr 20  &   4211.00317   &   4211.19962   &  {\it V\/}  &     300        &     51      \\
 14   &   2007 May 12  &   4231.92835   &   4232.34208   &  {\it V\/}  &     300        &     58      \\
 \hline
 \end{tabular}
 \medskip
\end{center}
\end{minipage}
\end{table*}

\begin{table*}
\centering
\begin{minipage}{175mm}
\begin{center}
  \bf Table 2. \rm Sine curve fits, times of maximun optical light and {\it RXTE} ASM X-ray data  for 4U 1735--44. \\
  \vspace*{0.125cm}
 \label{symbols}
 \begin{tabular}{@{}lcrccccc}
 \hline
  Obs. &  Cycle       &   HJD         &  Error   &       Amplitude        &      Mean $\Delta$      &   ASM Daily X-ray        &  No. ASM  \\
  No.  &  Number (N)  &   (-2450000)  &  (d)     &     {\it V\/} mag.     &       {\it V\/} mag.    &   Flux Units $cs^{-1}$   &  Dwells   \\
 \hline
   1   &     24547    &    662.0435   &  0.0046  &   0.188  $\pm$  0.028  &   1.190  $\pm$  0.010   &     8.97  $\pm$  0.50    &     0     \\
   2   &     24691    &    689.9535   &  0.0018  &   0.254  $\pm$  0.017  &   1.335  $\pm$  0.006   &    12.89  $\pm$  0.28    &     8     \\
   3   &     28550    &   1437.9459   &  0.0021  &   0.236  $\pm$  0.018  &   1.345  $\pm$  0.006   &    10.81  $\pm$  0.43    &     7     \\
   4   &     28571    &   1442.0236   &  0.0032  &   0.180  $\pm$  0.016  &   1.375  $\pm$  0.007   &    13.25  $\pm$  0.23    &    10     \\
   5   &     29851    &   1690.1214   &  0.0046  &   0.116  $\pm$  0.016  &   1.421  $\pm$  0.006   &    12.97  $\pm$  0.42    &     9     \\
   6   &     31935    &   2094.0848   &  0.0030  &   0.183  $\pm$  0.018  &   1.499  $\pm$  0.006   &    10.98  $\pm$  2.48    &     1     \\
   7   &     31940    &   2095.0451   &  0.0030  &   0.145  $\pm$  0.014  &   1.216  $\pm$  0.005   &    12.35  $\pm$  0.50    &     6     \\

   8   &     37296    &   3133.2238   &  0.0022  &   0.193  $\pm$  0.012  &   1.152  $\pm$  0.005   &    17.85  $\pm$  0.47    &     0     \\
  $9^{ a}$   &        &               &          &                        &   1.040  $\pm$  0.007   &    21.34  $\pm$  0.33    &    18     \\
  10   &     37337    &   3141.1581   &  0.0019  &   0.282  $\pm$  0.015  &   1.202  $\pm$  0.006   &    18.89  $\pm$  0.37    &     2     \\
  11   &     39204    &   3503.0260   &  0.0058  &   0.114  $\pm$  0.022  &   1.183  $\pm$  0.008   &    17.25  $\pm$  0.43    &     0     \\
  12   &     39209    &   3504.0359   &  0.0022  &   0.217  $\pm$  0.016  &   0.964  $\pm$  0.005   &    16.76  $\pm$  0.41    &     8     \\
  13   &     42857    &   4211.1238   &  0.0033  &   0.253  $\pm$  0.027  &   1.299  $\pm$  0.010   &    14.64  $\pm$  0.51    &     3     \\
  14   &     42965    &   4232.0368   &  0.0045  &   0.179  $\pm$  0.027  &   1.314  $\pm$  0.011   &    12.94  $\pm$  0.39    &    13     \\
 \hline
 \end{tabular}
 \medskip
\end{center}
\hspace*{3.0cm} {a}{ Observation not suitable for a sine curve fit. }  \\
\end{minipage}
\end{table*}

The optical counterpart of 4U 1735--44, V926 Scorpii, has been observed 
on a number of occasions since its identification in 1977 by McClintock 
et al. (1977). A collection of all these photometric data, covering 
the interval from 1984 July to 1993 August, was compiled by Augusteijn 
et al. (1998) (see references therein) to derive an accurate photometric 
ephemeris. Detailed spectroscopic studies have been reported by Smale \& 
Corbet (1991), Augusteijn et al. (1998) and Casares et al. (2006). 
This latter observation of 4U 1735--44 (and 4U 1636--53) is particularly 
relevant for the latter discussion as it established an important 
spectroscopic ephemeris, effectively defining a dynamical phase zero, 
by observing emission from the optical donor star. 

The paper is organised as follows. In Section 2 we report new 
photometric light curves of 4U 1735--44 obtained over the period 
1997 August to 2007 May. In Section 3 we use our new optical data 
to revise the Augusteijn et al. (1998) ephemeris and in Section 4 
we explore aspects of the historical X-ray data to look for possible 
correlated changes. In Section 5 we briefly discuss the 4U 1735--44 system 
based on our new observations.

\section{Optical Observations}
All the optical observations were made between 1997 and 2007 using the 
1-m telescope at the Mt. Canopus Observatory, University of Tasmania. 
The observations used standard Johnson {\it V\/} and {\it I\/} filters and 
the CCD reduction procedure was identical to that described in Giles, Hill 
\& Greenhill (1999). All times presented in this paper have been corrected 
to Heliocentric Julian Dates (HJD) and a complete journal of the 
observations is given in Table 1. Throughout this paper, except 
when discussing spectroscopic observations, phase zero 
is defined as superior conjunction of the companion star (neutron star 
closest to the Earth) when the system optical flux is at a maximum.

\begin{figure*}
\noindent
\begin{minipage}[b]{0.48\linewidth}
  \centering\epsfig{file=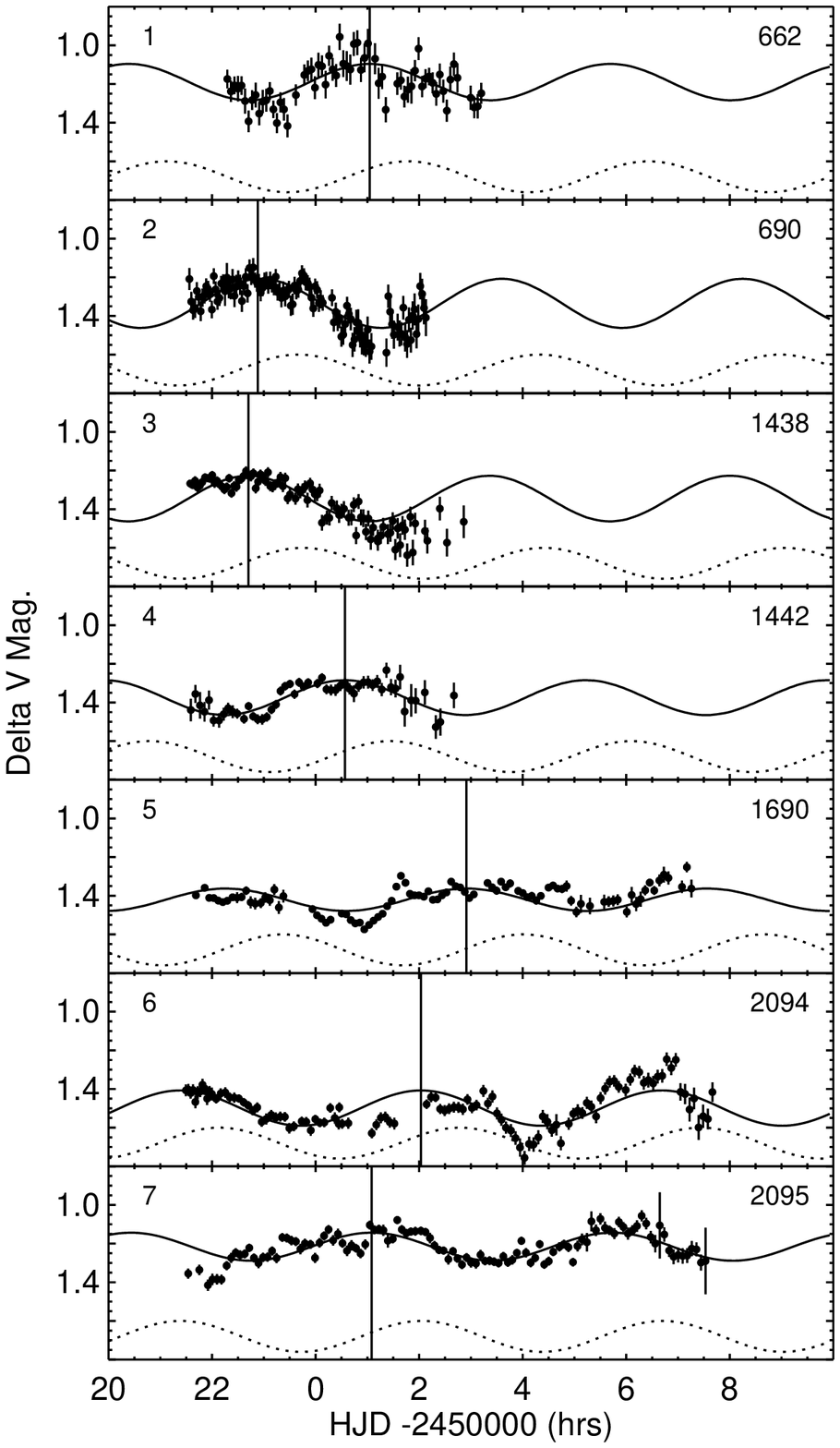, width=8.6cm }
\end{minipage}\hfill
\begin{minipage}[b]{0.48\linewidth}
  \centering\epsfig{file=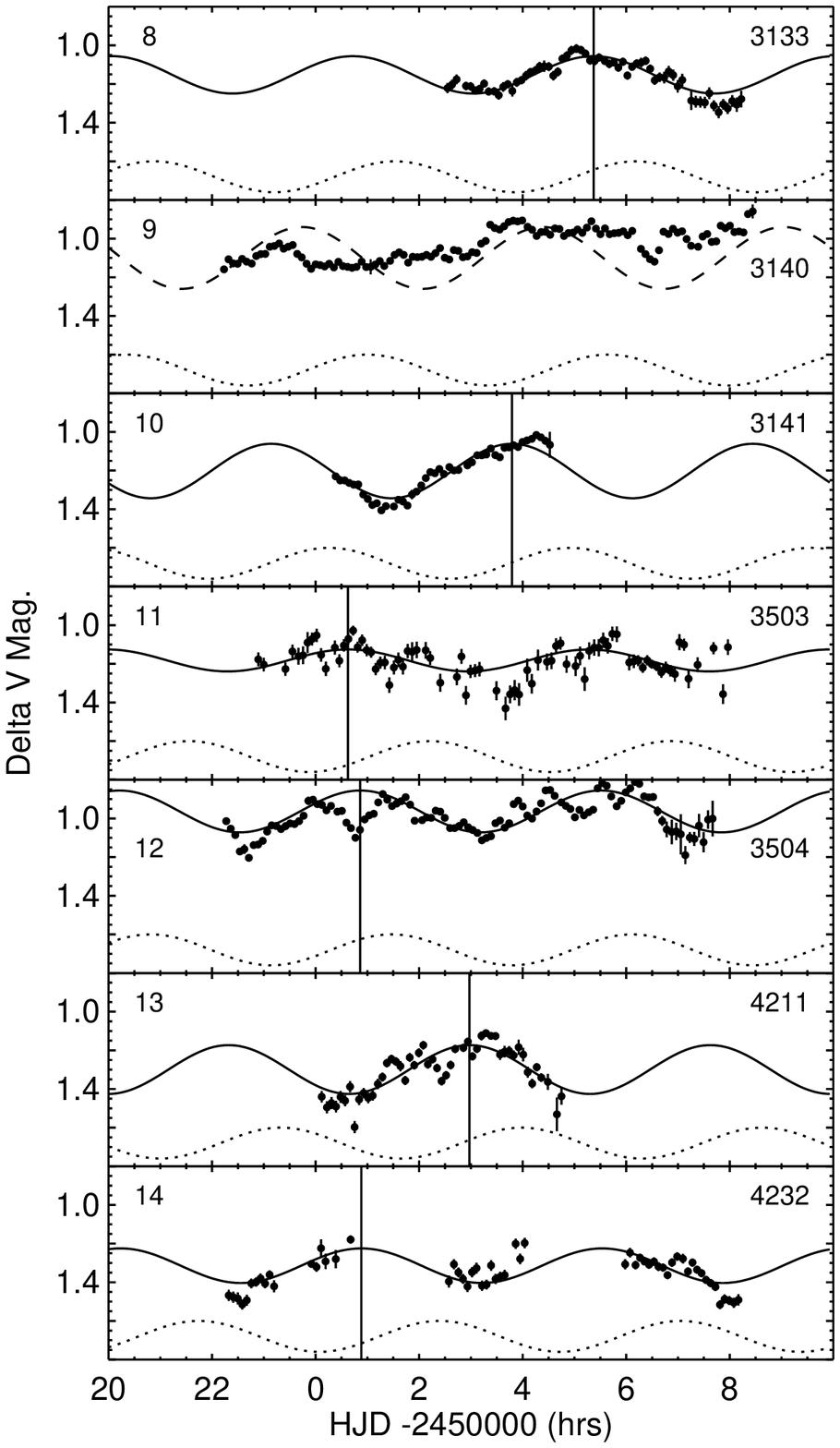, width=8.6cm }
\end{minipage}
\caption{The {\it V\/}-band light curves for 4U 1735--44. The solid traces 
through the data points mark our best fit sine curves to each nights 
observations. The dashed curves show the ephemeris predictions of Augusteijn 
et al. (1998) for the same nights with an arbitrary offset and 
amplitude. The number to the right in each panel refers to the HJD 
starting at zero hours within each light curve. The number to the left 
in each panel refers to the observation number listed in Table 2. 
Vertical lines mark the cycle count epochs listed in Table 2. 
The dashed trace for observation 9 is derived from the new ephemeris 
and is not a fit to the data points in that panel. }
\end{figure*}

All the images were reduced with between 3--7 nearby stars being used 
as local standards. However, the light curves in Fig. 1  plot the 
differential magnitude between 4U 1735--44 and only one of these stars 
which can be located on the finder chart in Jernigan et al. (1977). 
This brighter secondary standard is star number 6 on their 2S1735--444 
chart which we find to have a {\it V\/} magnitude of 16.10 $\pm$ 0.02. 
4U 1735--44 is star number 5 on this same chart and is $\sim1.4$ mag fainter. 
For the 1997 observations the telescope was equipped with an SBIG CCD 
camera having 375 x 242 pixels with an image scale of 0.42 $\times$ 0.49 
arcsec pixel$^{-1}$. On the nights of 1997 August 1 and 29 continuous 
pairs of {\it V\/} and {\it I\/} integrations were obtained but 
the {\it I\/}-band data are not discussed further in this paper. 
For the 1999 and later observations the telescope was equipped with 
a SITe CCD camera having 512 x 512 pixels with an image scale of 
0.42 arcsec pixel$^{-1}$. 

The entire data set represented in Tables 1 and 2 are plotted in Fig. 1. 
Observation 9  has no clear modulation and has therefore not been fitted 
with a sine curve. The data are included since the mean level is used 
in section 4 of this paper and it also illustrates this unmodulated state 
at the time of highest X-ray flux. Observation 12 is also interesting 
as it appears to show two cycles containing a definite dip, midway 
in phase between maximum brightness.

\section{Optical Maximum Ephemeris}

The ephemeris for times of maximum optical light given by Augusteijn et al. 
(1998) was HJD = 2447288.0143(25) + [ N $\times$ 0.19383351(32) ] 
where the errors are indicated in the round brackets ($\pm \sigma$) and 
N is the cycle number starting from zero. This ephemeris was based on 
observations made between 1984 July 22 and 1993 July 28 and covered a total 
of 16,986 binary periods. We have fitted a sine curve to each new night's 
observations listed in Table 1 taking the amplitude, phase and mean as free 
parameters but fixing the binary period at the value given by Augusteijn 
et al. (1998). The appropriate sine curve fits are shown as solid traces 
through the respective data points in Fig. 1. From these fits we 
derive the times of optical maxima, peak-to-peak amplitudes and mean 
intensities listed in Table 2. This table is intended to be complimentary 
to the similar table 3 of Augusteijn et al. (1998) and continues 
the same cycle number sequence. 

The predictions from the previous Augusteijn et al (1998) ephemeris 
are also shown on Fig. 1 as the dotted traces in the lower part 
of each light curve panel. There is a significant phase shift evident 
between the two sets of sine curves. The new observed maxima also fall 
increasingly earlier than expected, indicating a different period, over 
the additional $\sim$25,979 binary periods to 2007 May. However, a quick 
examination shows that the earlier period was close enough to avoid any 
cycle count ambiguity through the new data sets and, as stated earler, 
the previous cycle sequence is maintained in Table 2.

\begin{figure}
  \epsfig{file=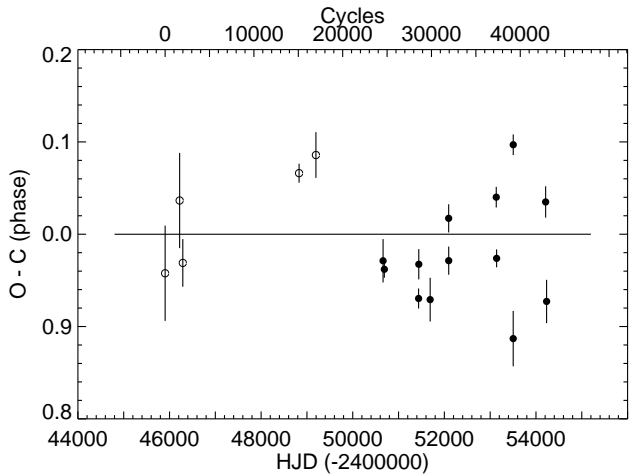, width=8.6cm } 
  \caption{The {\it O\/} - {\it C\/} times of maximum optical 
light for 4U 1735--44 plotted against time for the best fit linear 
model. The open circles denote the earlier points from Augusteijn et al. 
(1998) and the filled circles are the new optical data.}
\end{figure}

We then used our new data, together with the data in table 3 of 
Augusteijn et al (1998,) to perform the usual {\it O\/} - {\it C\/} 
analysis on the entire data set and the results are plotted in Fig. 2. 
Our new linear fit for the time of maximum optical light has the ephemeris 
HJD = 2445904.0494(90) + [ N $\times$ 0.19383222(29) ] 
with a value of $\chi^2 = 253.5$ for 16 dof and a mean phase scatter about 
phase zero of 0.061. The relatively poor $\chi^2$ value reflects 
the intrinsic scatter of the data points rather than their error bars. 
Deriving a highly accurate and reliable 
ephemeris for systems like 4U 1735--44 (and 4U 1636--53) is not simple 
due to the lack of any sharp eclipse type feature in the optical light 
curve and the fact that the profile is quite variable from cycle to cycle. 
For these reasoms and the `noise' in the scatter of the {\it O\/} - {\it C\/}
values in Fig. 2. there is still no significant period derivative term. 
An upper limit for the $\dot P$ term gives 
$\mid P/\dot P\mid$ $\lesssim 6.5 \times$10$^{6}$ yr. 

Until relatively recently it had proved impossible to reliably identify 
features in the optical spectrum of 4U 1735--44 that originated from the 
irradiated donor companion star which could be used to define a dynamical 
phase zero. This has been achieved for a number of LMXB systems by following 
the Bowen transition feature and Casares et al. (2006) have obtained a 
spectroscopic based ephemeris for 4U 1735--44 using observations made in 
2003 June. Phase zero, in this spectral definition, corresponds to inferior 
conjunction of the companion (donor) star and occurs at
HJD = 2452813.495(3) + [ N $\times$ 0.19383351(32) ] where the period was 
taken as the Augusteijn et al. (1998) value. Note that this definition 
has a 0.5 phase shift with respect to that used for optical intensity. 
Our new photometric ephemeris places the optical maximun at a phase 
of 0.47 $\pm$ 0.05 on this spectral ephemeris. 

Since the spectral ephemeris date falls well within the sequence of 
new optical observations (see Fig. 3) the phase zero error in our new 
optical ephemeris is primarily due to the uncertainty in defining when 
the optical maximum occurs rather than to period uncertainty accumulations 
or the spectral phase zero error.

\begin{figure}
  \epsfig{file=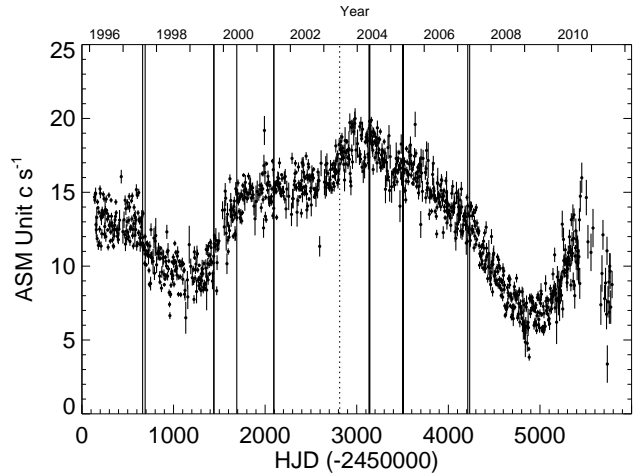, width=8.6cm }
  \caption{The ASM light curve for 4U 1735--44 from 1996 February 21, just 
after the launch of {\it RXTE}, to 2011 August 30. The ASM data (1.5--12 keV) 
were extracted as one day values and then averaged over a four day interval. 
Solid  vertical lines mark the dates of the optical observations listed in 
Table 1. The dotted vertical line marks the date of the spectral phase zero 
observation of Casares et al. (2006).}
\end{figure}

\section{Optical - X-ray correlations}

Since our data set spans an $\sim 10$ year interval and overlaps with much 
of the duration of the {\it RXTE} mission we have also examined the All Sky 
Monitor (ASM) (Levine et al (1996)) database to look at the X-ray light 
curve history for 4U 1735--44. Fig. 3 shows this data together with the 
dates on which our optical observations were made.

We have attempted to extract the ASM fluxes corresponding to the exact times of 
the observations listed in Table 1 in order to compare them to the corresponding 
optical intensities and amplitude modulations listed in Table 2. 
ASM X-ray fluxes can be obtained as individual dwell cycle values, effectively 90 s
integrations, or as daily averages which are summations of all the dwell cycles 
falling within 24 h intervals. Unfortunately, due to a combination of orbital and 
aspect constraints, the dwell cycle values for  4U 1735--44 occur in irregularly 
spread groups with gaps containing no observations. Although daily values are virtually 
always available they may be calculated from data within 24 h intervals that fall 
outside the precise span of any particular stretch of optical observations. 
Fig. 1 shows that the durations of the optical observations vary between 
4--10 h and Table 2 shows that the number of dwell cycles within each optical 
sequence varies from 0--18. Table 2 also shows the nearest, or encompassing, 
daily ASM X-ray flux intensities (bands A+B+C, 1.5--12 keV). 

For the analysis we have chosen to use daily values as many optical nights
have few or no dwell cycles. A visual inspection of the appropriate ASM 
data sections suggests that, in general, the daily values represent reasonable 
local estimates even where they do not overlap with the optical data. Any effort 
to be more selective rapidy becomes problematic and rather arbitary in nature. 
Fig. 4 shows that there is a roughly linear trend between the X-ray flux 
and the optical intensity with the optical source brightening, as expected, 
when the X-ray flux increases. The linear fit, to all the observations, has a 
value of $\chi^2 = 93.9$ for 12 dof and a modest correlation coefficient of 0.64.  
The slope of the line gives a $\Delta$ optical increase of 59 percent (0.5 mag.) 
for a $\Delta$ X-ray increase of 69 percent. This suggests that the non-X-ray 
induced optical flux from the companion is  $\lesssim 14$ percent of the 
total light from the system as expected from a late type dwarf star.

\begin{figure}
  \epsfig{file=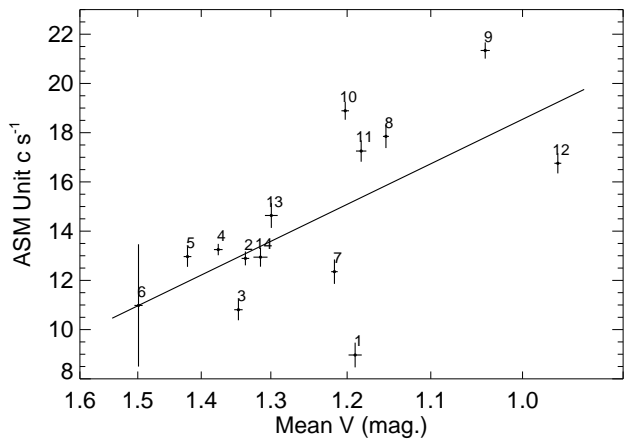, width=8.6cm }
  \caption{The {\it RXTE} ASM one-day X-ray intensity plotted against mean optical 
brightness for the data in Table 2. The X axis plots the difference to the 
brighter comparison star so lower values (points to the right) are brighter. 
The numbers indicate the observation sequence listed in Table 2. }
\end{figure}

In contrast, there is no apparent correlation between the X-ray flux and the 
amplitude modulation for 4U 1735--44 in Fig. 5 though the modulation error 
bars are relatively large in this case. Perhaps only observation 1 seems 
to stand outside the general linear trend in Fig. 4 and this happens 
to be both the oldest optical data and also has no contained ASM dwell 
cycles. However, the daily X-ray flux value appears to be consistent 
with that for adjacent days and dwell cycles and the relative values 
of the various calibration stars on this night are also consistent 
with the many later CCD sequences. 

The entire ASM data set has been folded, 
modulo the period, using the new ephemeris and there is no evidence for any 
orbital modulation, dips or eclipses. This applies for combined energy bands 
(A+B+C) as well as just the low (A, 1.5--3 keV) and high (C, 5--12 keV) bands. 
This is also true whether the data is selected for when the source is bright 
(above the mean for the whole record) or weak (below the mean).

\begin{figure}
  \epsfig{file=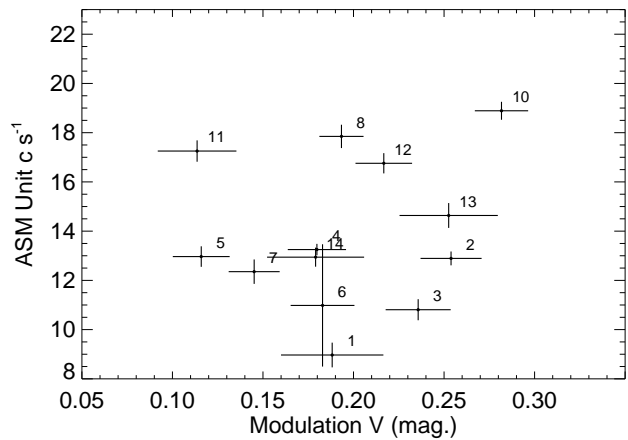, width=8.6cm }
  \caption{The {\it RXTE} ASM one-day X-ray intensity plotted against optical 
modulation amplitude (peak-to-peak) for the data in Table 2. No obvious trend 
is apparent in this data. Observation 9 is missing from this plot. }
\end{figure}

\section{Discussion}

In the traditional model of an LMXB system there are three regions which 
contribute to its optical variability due to X-ray heating. These are the 
accretion  disc itself, a bright hot spot or bulge on the outer edge of the 
accretion disc due to the impact of inflowing material and the hemisphere of 
the companion facing the neutron star which is not shadowed by the accretion 
disc. For most LMXB systems the reprocessed X-ray optical flux is assumed to 
come from the facing hemisphere of the companion star and to dominate the 
optical light from the rest of the system (van Paradijs 1983, van Paradijs \& 
McClintock 1995). The optical maximum was therefore assumed to occur when 
the companion was on the far side of the neutron star but variations about 
the mean modulation profile were expected due to gas flows causing 
various X-ray shielding effects (Pedersen et al. 1982). Since the 
magnetosphere of the neutron star in the  4U 1735--44 system is so small 
the Roche lobe overflow through the inner Lagrange point (L1) must form a 
standard Keplerian accretion disc (Frank, King \& Raine, 1992). 

Smale \& Mukai (1988) investigated the optical flux and modulation created 
by the persistent X-ray heating of the companion's facing hemisphere, and 
although they assumed this scenario in their modelling they commented that 
a thick disc would also likely contribute to the variable optical flux. 

We have not attempted any detailed modelling of our observations but a 
number of comments can be made. The earlier optical ephemeris due to 
Augusteijn  et al. (1998) placed maximum optical light for 4U 1735--44 
at phase 0.68 $\pm$ 0.06 on the spectral ephemeris of Casares  et al. 
(2006) which was between the expected maximum visibility of the 
irradiated donor and the disc bulge. This was in contrast to 
the result for 4U 1636--53 where their spectral ephemeris and the 
optical ephemeris of Giles et al. (2002) placed optical maximum 
at a more reasonable phase of 0.47 $\pm$ 0.06. Our new photometric 
ephemeris for 4U 1735--44 resolves this issue by placing the optical 
maximum much closer to phase 0.5 (at 0.47 $\pm$ 0.05). 
Indeed, Casares  et al. (2006) comment on the uncertainty of the 
Augusteijn et al. (1998) ephemeris and Fig. 2 clearly shows how an 
ephemeris based on only the 5 earliest data points leads to a 
significant phase error of $\sim 0.2$ after the passage of 10-yr. 

The correlation between optical flux and X-ray flux, evident 
in Fig. 4, is not unexpected, but Fig. 5 shows no apparent correlation
between optical modulation amplitude and X-ray flux. 
If the X-ray induced optical flux were sitting on a substantial 
and constant base component from the companion star then Fig. 5 would be 
expected to show a small  increase in modulation percentage (essentially 
just $\Delta$ mag. for small values $\lesssim 0.30$) for increasing X-ray 
flux. There is no evidence for this. Therefore, the companion contribution 
must be minor and may be obscured, on Fig. 5, by the variability in 
the relationships between the different components generating, or 
obscuring, the optical flux. The X axis in Fig. 5 is still in 
magnitudes, rather than on a linear intensity scale, as in Fig. 4. 

The X-ray flux and optical intensity were also shown to be positively  
correlated in observations of 4U 1636--53 by Shih, Charles \& Cornelisse (2011) 
who performed a 100 day, monitoring program of this source in which daily 
spot optical measurements were made. Only the lower energy X-ray 
intensity (1.5--12 keV {\it RXTE} ASM) showed a positive correlation, the 
high energy X-ray flux (15--50 keV {\it Swift} BAT) being 
anti-correlated. In the present paper, the optical and X-ray data 
have been measured across binary cycles and many hours respectively 
and thus hopefully represent more reasonable `steady-state' average values. 
There is insufficient data to be certain, but there is a suggestion in the 
light curves that at the very highest X-ray flux the optical modulation 
has almost gone away and the faster optical variability decreased. 
Perhaps this is an indication of an optical state change. The continuing 
lack of any orbital modulation signature in the folded X-ray data suggests 
$i \le 60^{\circ}$, based on the modelling of Frank, King \& Lasota (1987). 

Several key aspects of the standard model for an LMXB system are 
supported by the following results: \vspace*{0.2cm} \\
\hspace*{0.2cm} $\bullet$ Optical maximum at spectral phase 0.47 $\pm$ 0.05  \\
\hspace*{0.2cm} $\bullet$ Positive correlation, optical flux -- X-ray flux  \\
\hspace*{0.2cm} $\bullet$ No correlation, optical amplitude -- X-ray flux  \\
\hspace*{0.2cm} $\bullet$ Non-X-ray induced optical flux is $\lesssim 14$ percent  \\
\vspace*{0.1cm}

The dynamical ephemeris for 4U 1735--44 is now obtained by combining the 
new period derived in this paper with the phase zero epoch from 
Casares et al. (2006). Phase zero, in this spectral definition, again 
corresponds to inferior conjunction of the companion star. The ephemeris 
is therefore HJD = 2452813.495(3) + [ N $\times$ 0.19383222(29) ].

Clearly, on-going occasional optical photometry and spectroscopy are 
desirable for both 4U 1735--44 and 4U 1636--53 to monitor any possible changes, 
in order to better understand LMXB systems in general. These two X-ray 
binaries continue to be quite similar and are apparently viewed from much 
the same perspective. Future simultaneous X-ray and optical studies may be 
problematic with the recent decommissioning of the long running {\it RXTE} 
mission in 2012 January. A number of alternative, somewhat equivalent, 
ASM type systems are presently operating. Perhaps the {\it LOFT} mission, 
if selected for flight by ESA, will provide a more continuous X-ray coverage.

\section{Acknowledgements}
We thank John Greenhill for assistance at the Mt. Canopus Observatory 
during the early part of this work and Stefan Dieters for helpful 
comments. ABG thanks the ACE CRC at the University of Tasmania 
for the use of computer facilities. The Mt. Canopus Observatory 
received some financial support from David Warren.

\bsp 

\label{lastpage}

\end{document}